\documentclass[
aip,
final,
reprint,
floatfix,
]{revtex4-2}

\usepackage[dvips]{graphicx}
\usepackage[utf8]{inputenc}
\usepackage{amsmath,amssymb,booktabs,latexsym,MnSymbol,bm}
\usepackage{setspace}

\usepackage[%
  colorlinks=true,
  urlcolor=blue,
  linkcolor=blue,
  citecolor=blue,
  allcolors=blue
]{hyperref}

\begin{document}

\setlength{\parskip}{0pt}

\title{Characterizing unstructured data with the nearest neighbor permutation entropy}

\author{Leonardo G. J. M. Voltarelli} 
\affiliation{Departamento de F\'isica, Universidade Estadual de Maring\'a -- Maring\'a, PR 87020-900, Brazil}

\author{Arthur~A.~B.~Pessa} 
\affiliation{Departamento de F\'isica, Universidade Estadual de Maring\'a -- Maring\'a, PR 87020-900, Brazil}

\author{Luciano Zunino} 
\affiliation{Centro de Investigaciones \'Opticas (CONICET La Plata - CIC - UNLP), C.C. 3, 1897 Gonnet, La Plata, Argentina}
\affiliation{Departamento de Ciencias B\'asicas, Facultad de Ingenier\'ia, Universidad Nacional de La Plata (UNLP), 1900 La Plata, Argentina}

\author{Rafael~S.~Zola} 
\affiliation{Departamento de F\'isica, Universidade Estadual de Maring\'a -- Maring\'a, PR 87020-900, Brazil}
\affiliation{Departamento de F\'isica, Universidade Tecnol\'ogica Federal do Paran\'a, Apucarana, PR 86812-460, Brazil}

\author{Ervin K. Lenzi} 
\affiliation{Departamento de F\'isica, Universidade Estadual de Ponta Grossa -- Ponta Grossa, PR 84030-900, Brazil}

\author{Matja{\v z}~Perc} 
\affiliation{Faculty of Natural Sciences and Mathematics, University of Maribor, Koro{\v s}ka cesta 160, 2000 Maribor, Slovenia}
\affiliation{Community Healthcare Center Dr. Adolf Drolc Maribor, Vo{\v s}njakova ulica 2, 2000 Maribor, Slovenia}
\affiliation{Complexity Science Hub Vienna, Josefst{\"a}dterstra{\ss}e 39, 1080 Vienna, Austria}
\affiliation{Department of Physics, Kyung Hee University, 26 Kyungheedae-ro, Dongdaemun-gu, Seoul, Republic of Korea}

\author{Haroldo V. Ribeiro} 
\email{hvr@dfi.uem.br}
\affiliation{Departamento de F\'isica, Universidade Estadual de Maring\'a -- Maring\'a, PR 87020-900, Brazil}
\date{\today}

\date{\today}

\begin{abstract}
Permutation entropy and its associated frameworks are remarkable examples of physics-inspired techniques adept at processing complex and extensive datasets. Despite substantial progress in developing and applying these tools, their use has been predominantly limited to structured datasets such as time series or images. Here, we introduce the $k$-nearest neighbor permutation entropy, an innovative extension of the permutation entropy tailored for unstructured data, irrespective of their spatial or temporal configuration and dimensionality. Our approach builds upon nearest neighbor graphs to establish neighborhood relations and uses random walks to extract ordinal patterns and their distribution, thereby defining the $k$-nearest neighbor permutation entropy. This tool not only adeptly identifies variations in patterns of unstructured data, but also does so with a precision that significantly surpasses conventional measures such as spatial autocorrelation. Additionally, it provides a natural approach for incorporating amplitude information and time gaps when analyzing time series or images, thus significantly enhancing its noise resilience and predictive capabilities compared to the usual permutation entropy. Our research substantially expands the applicability of ordinal methods to more general data types, opening promising research avenues for extending the permutation entropy toolkit for unstructured data.
\end{abstract}

\maketitle

\begin{quotation}
\small Our work contributes to the pressing need for methodologies capable of efficiently handling large-scale unstructured data by introducing the \textit{$k$-nearest neighbor permutation entropy}. This tool extends the fundamental premise of investigating the relative ordering of time series elements inaugurated by permutation entropy to non-gridded datasets such as point clouds, point processes, and geospatial data while keeping the simplicity and interpretability of the original method. The $k$-nearest neighbor permutation entropy adeptly identifies variations in patterns of unstructured data with a level of precision markedly superior to celebrated complex system methodologies. Our approach not only expands the application horizon of permutation entropy but also demonstrates enhanced robustness and discrimination capabilities in analyzing structured data such as time series and images.
\end{quotation}

\section{Introduction}

Scientific fields are universally witnessing an unprecedented surge in the volume and intricacy of digital data available for research. This data revolution~\cite{mattmann2013vision, world2021world} has spurred an urgent demand for methods that are simultaneously simple, interpretable, robust, computationally efficient, and yet capable of extracting valuable information from large-scale databases. Leveraging a long-standing tradition of uncovering fundamental principles from complex systems, physics-inspired methods have emerged as remarkably effective for navigating these challenges. A notable example is the permutation entropy~\cite{bandt2002permutation} and its underlying idea positing that the relative ordering of data points is instrumental for system characterization. This method not only fulfills the requirements for managing complex and voluminous data but also offers the added benefit of facilitating reproducible results. Permutation entropy has found successful applications related to data analysis across various disciplines, including engineering~\cite{yan2012permutation}, medical sciences~\cite{nicolaou2012detection, santos2022characterisation}, econophysics~\cite{zunino2009forbidden}, climate sciences~\cite{garland2018anomaly}, aerodynamics~\cite{pessa2022clustering}, condensed matter physics~\cite{sigaki2019estimating, pessa2022determining}, and even visual arts~\cite{sigaki2018history}, evidencing its versatility and utility in diverse research domains~\cite{zanin2012permutation, riedl2013practical, amigo2015ordinal, keller2017permutation, pessa2021ordpy, amigo2023ordinal}.

The success of permutation entropy extends beyond its practical applications, as its core principle of deriving a probability distribution from ordinal patterns in data has served as a foundational framework for a multitude of data analysis tools. These advancements include the calculation of other quantifiers from the ordinal probability distribution, such as conditional entropy~\cite{unakafov2014conditional}, complexity-entropy plane~\cite{rosso2007distinguishing}, and complexity-entropy curves~\cite{ribeiro2017characterizing}. It also encompasses the analysis of forbidden patterns~\cite{amigo2006order, amigo2007true}, the treatment of equal values patterns~\cite{bian2012modified, zunino2017permutation, cuesta2018patterns}, and the exploration of multiscale ordinal patterns~\cite{zunino2010permutation, zunino2012distinguishing}. The original methodology has further inspired the creation of ordinal networks~\cite{small2013complex, mccullough2015time, mccullough2017multiscale, zhang2017constructing, small2018ordinal, pessa2019characterizing, borges2019learning, pessa2020mapping}, which explore transitions between ordinal patterns, thereby unveiling novel perspectives on the dynamics of complex systems. Additionally, the methodology underlying permutation entropy has been extended to accommodate two-dimensional data~\cite{ribeiro2012complexity, zunino2016discriminating, bandt2023two}, significantly broadening its applicability to the analysis of image datasets.

Despite substantial theoretical progress and successful application in data analysis, the permutation entropy framework is predominantly limited to gridded data, with only one recent exception related to its usage in networked signals~\cite{fabila2022permutation}. This constraint stems from the initial conceptualization of permutation entropy, which focused on discerning ordering patterns among sequential observations within time series~\cite{bandt2002permutation}. The extension to two-dimensional data similarly inherits this limitation, concentrating on ordinal patterns derived from rectilinear subarrays~\cite{ribeiro2012complexity}. Moreover, even with multiscale generalizations~\cite{zunino2010permutation, zunino2016discriminating}, the analysis still relies on gridded partitions composed of evenly spaced elements, and the idea of multiple scales refers to the use of different time lags for sampling time series~\cite{zunino2010permutation} or adjusting resolution levels in image analysis~\cite{zunino2016discriminating}. Overcoming this limitation has the potential to significantly expand the utility of permutation entropy to a wider array of data types, including point cloud data (such as three-dimensional scans from LiDAR sensors or medical imaging), point process data (such as spatial distribution of disease outbreaks or earthquake occurrences), geospatial data (such as GPS tracking of animals, city indicators in urban systems, or sampling points related to environmental monitoring), and various other data structures that are not uniformly sampled in either time or space. Furthermore, tackling this challenge opens up promising research avenues related to extending and adapting the extensive toolkit of the permutation entropy framework to these various data structures. 

Here, we introduce the nearest neighbor permutation entropy, an innovative generalization of the permutation entropy designed to accommodate unstructured data across multidimensional spaces. Our method leverages a nearest neighbor graph to establish neighborhood relations among data points and uses random walks on this graph to generate time series. These time series allow the extraction of ordinal patterns, calculation of their probability distribution, and estimation of Shannon entropy, thereby defining the nearest neighbor permutation entropy of general data structures irrespective of their spatial or temporal configuration and dimensionality. We investigate the efficacy of this new technique in analyzing spatial patterns, revealing that it not only adeptly identifies variations in these patterns, but also does so with a level of precision that significantly surpasses conventional measures such as spatial autocorrelation. We further show that our approach improves the characterization of irregularly sampled time series by intuitively incorporating information about time gaps. Beyond broadening the scope of the standard approach, we demonstrate that the nearest neighbor permutation entropy is equally applicable to regular time series and images, but offers enhanced robustness against noise and improved discrimination capability compared to the conventional permutation entropy.

\section{Nearest neighbor permutation entropy}

\begin{figure*}[ht!]
  \centering
  \includegraphics[width=1\textwidth, keepaspectratio]{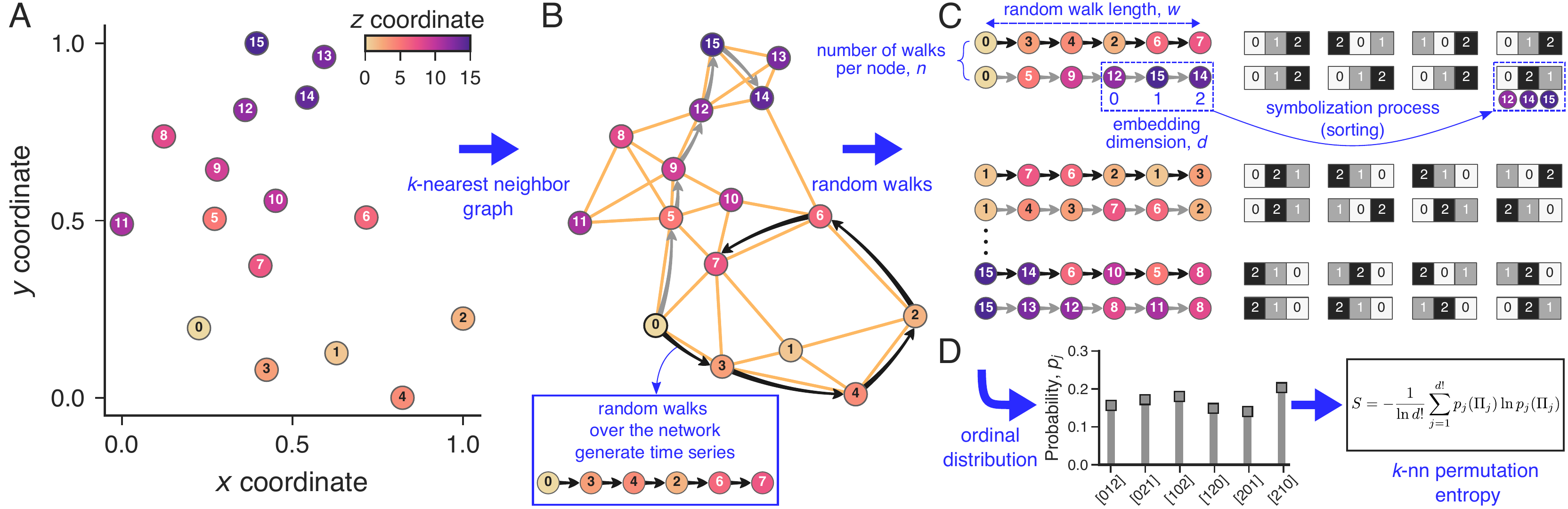}
  \caption{Implementation of the $k$-nearest neighbor permutation entropy. (A) Illustration of a dataset with irregularly distributed data points $\{z_i\}_{i=1,\dots,N}$ in the $xy$-plane where each coordinate pair $\vec{r}_i = (x_i,y_i)$ is associated with a value $z_i$. (B) Initially, we construct a $k$-nearest neighbor graph using the data coordinates to define neighborhood relationships. In this graph, each data point $z_i$ represents a node, with undirected edges connecting pairs $i\leftrightarrow j$ when $j$ is among the $k$-nearest neighbors of $i$ ($k=3$ in this example). (C) Subsequently, we execute $n$ biased random walks of length $w$ starting from each node, sampling the data points to generate time series ($n=2$ and $w=6$ in this example). We then apply the Bandt-Pompe approach to each of these time series. This involves creating overlapping partitions of length $d$ (embedding dimension) and arranging the partition indices in ascending order of their values to determine the sorting permutations for each partition ($d=3$ in this example). (D) Finally, we evaluate the probability of each of the $d!$ possible permutations (ordinal distribution) and calculate its Shannon entropy, thereby defining the $k$-nearest neighbor permutation entropy.
  }
\label{fig:1}
\end{figure*}

We begin by examining data points within a potential multidimensional space. These points are characterized by a position vector $\vec{r}_i$ and an associated value $z_i$, for $i=1, 2, \dots, N$, where $N$ is the dataset size. For example, two-dimensional position vectors $\vec{r}_i = (x_i, y_i)$ may denote the spatial locations of cities, while the values $z_i$ could represent urban indicators such as population or gross domestic product. The initial step in our method involves constructing a $k$-nearest neighbor graph based on the position vectors $\vec{r}_i$. In this graph, nodes correspond to individual data points, and undirected edges connect nodes to their $k$-nearest neighbors based on a chosen distance metric. Although any distance metric is applicable, all our analyses use the Euclidean distance. Figure~\ref{fig:1}A depicts a hypothetical dataset of $N=16$ points irregularly arranged on the $xy$-plane, and Figure~\ref{fig:1}B illustrates the resulting nearest neighbor graph obtained for this dataset with $k=3$. 

This graph representation is fundamental for establishing neighborhood relations among the unstructured data points, which are explored in the second step of our approach. In this step, we sample the $z_i$ values by initiating $n$ random walks of length $w$ from every node in the graph. These random walkers are intentionally biased to produce a depth-first sampling of the $z_i$ values, aiming to visit nodes increasingly distant from the source node. Specifically, we use a strategy akin to \textit{node2vec}~\cite{grover2016node2vec}, an algorithm designed to produce vector representations of network nodes. The process is a second-order random walk, wherein the walker decides to move from its current position $b$ to a subsequent position $c$, considering its previous position $a$, according to the unnormalized transition probability
\begin{equation}
\rho_{bc}  = \begin{cases}
    1/\lambda & \text{if}~s_{ac}=0\\
    1 & \text{if}~s_{ac}=1\\
    1/\beta & \text{if}~s_{ac}=2
\end{cases}\,,
\end{equation}
where $s_{ac}$ denotes the shortest-path distance between nodes $a$ (previous position) and $c$ (subsequent position), while $\lambda$ and $\beta$ are positive parameters controlling the walker's bias. A high value of $\lambda$ exceeding 1 reduces the likelihood of the walker returning to its previous position $a$, while lower values favor backtracking movements. In turn, a high value of $\beta$ greater than 1 enhances the walker's preference toward moving to nodes near its prior position, whereas lower values make the walker more likely to move beyond the immediate neighborhood of its previous location. In all our applications, we set $\lambda=10$ and $\beta=0.001$ to ensure a depth-first sampling of the $z_i$ values. It is also worth noticing that, despite the existence of bias, the probabilistic nature of the process prevents walkers from becoming trapped by any dangling ends within the graph. Figure~\ref{fig:1}B illustrates two examples of such random walk starting from the node with $z_i=0$.

The random walks generate an ensemble comprising $N\times n$ trajectories of length $w$, where $N$ and $n$ denote the dataset size and the number of walks per node, respectively. These trajectories explore relations among adjacent data points and offer a natural way to define ordering patterns. Having these trajectories or time series, we can now apply the standard methodology proposed in the permutation entropy framework~\cite{bandt2002permutation}. We represent a given trajectory as $\{\tilde{z}_t\}_{t=1,\dots,w}$ and segment it into overlapping partitions 
\begin{equation}
    u_q = (\tilde{z}_q,\tilde{z}_{q+1},\dots,\tilde{z}_{q+d-1})\,,
\end{equation}
where $d$ is the partition length or embedding dimension, and $q=1,\dots,w-d+1$ indexes each partition. For each partition, we determine the permutation $\pi_q=(r_0, r_1, \dots, r_{d-1})$ of the index numbers $(0,1,\dots,d-1)$ that organizes the observations in $u_q$ in ascending order. Each of these permutations is defined by the inequality $\tilde{z}_{q+r_0} \leq \tilde{z}_{q+r_1} \leq \dots \leq \tilde{z}_{q+r_{d-1}}$, and in cases of equal values, their occurrence order is maintained (that is, if $\tilde{z}_{q+r_{s-1}}=\tilde{z}_{q+r_{s}}$ for some $s\in(1,\dots,d-1)$, then $r_{s-1}<r_s$). We replicate this procedure across all sampled trajectories, yielding $M=N \times n \times (w-d+1)$ permutation symbols or ordinal patterns. Figure~\ref{fig:1}C illustrates possible sampled trajectories and the process of obtaining permutation symbols. In this example, we have $n=2$ random walks of length $w=6$ per node and an embedding dimension $d=3$. For instance, the latest partition obtained from the second random walk starting from the node with value $z_i=0$ ($\{\tilde{z}_t\}_{t=1,\dots,6}=\{0,5,9,12,15,14\}$) is $u_4=(12,15,14)$; sorting its elements yields $12 \leq 14 \leq 15$ or $\tilde{z}_{4+0} \leq \tilde{z}_{4+2} \leq \tilde{z}_{4+1}$, and thus the corresponding ordinal pattern is $\pi_4=(0,2,1)$.

Furthermore, by analyzing all $M$ permutation symbols $\{\pi_s\}_{s=1,\dots,M}$ obtained from the sampled trajectories, we calculate the probability $p_j(\Pi_j)$ of each possible permutation symbol $\Pi_j$, where $j=1,\dots,d!$, by determining their relative frequency
\begin{equation}
    p_j(\Pi_j) = \frac{\text{total of permutations $\Pi_j$ in $\{\pi_s\}$}}{M}\,.
\end{equation}
The resulting probabilities constitute the ordinal probability distribution $P=\{\Pi_j\}_{j=1,\dots,d!}$ and its normalized Shannon entropy~\cite{shannon1948mathematical}
\begin{equation}
    S = -\frac{1}{\ln d!} \sum_{j=1}^{d!} p_j(\Pi_j) \ln p_j(\Pi_j)\,,
\end{equation}
defines our $k$-nearest neighbor permutation entropy (or $k$-nn permutation entropy, for short). Figure~\ref{fig:1}D illustrates this final step of our method. The values of $S$ quantify the degree of irregularity in the spread of data points $\{\vec{r}_i,z_i\}$.  We expect to find $S\approx0$ (lower limit) when adjacent values of $z_i$ exhibit a regular configuration characterized by the dominance of a single ordinal pattern. Conversely, $S\approx1$ (upper limit) suggests the absence of regular structures among adjacent values of $z_i$, indicating the lack of preference for specific ordinal patterns.

In addition to the embedding dimension $d$, which sets a scale for analyzing ordinal patterns, our approach introduces additional parameters for calculating the $k$-nn permutation entropy. The parameter $k$, representing the number of nearest neighbors, establishes the graph structure and controls the emphasis on local versus global data structures. An excessively small $k$ leads to a nearest neighbor graph with numerous small connected components, confining the sampled trajectories within these components and thus limiting the identification of patterns among distantly located data points. Conversely, an overly large $k$ results in a graph that resembles a fully connected structure, rendering the sampled trajectories comparable to a random selection of data points and driving entropy values towards the upper limit ($S\to1$). This interplay between local and global structure echoes the principles inherent in the uniform manifold approximation and projection (UMAP) method~\cite{mcinnes2018umap, mcinnes2018umapsoftware}, a leading-edge method for dimensionality reduction. UMAP constructs a weighted version of a $k$-nearest neighbor graph from high-dimensional data (termed a fuzzy simplicial complex~\cite{mcinnes2018umap}) and projects it into a lower-dimensional space using a force-directed graph layout algorithm. Similarly to our method, the choice of $k$ in UMAP essentially determines the extent to which local and global data structures are preserved in the low-dimensional projection. In our experiments with $N$ ranging from approximately $10^3$ to $10^5$ data points, we find that selecting a number of neighbors between $10$ and $50$ produces broadly similar results. Therefore, we opt to report results for $k=25$ unless otherwise specified. In general, the selection of $k$ might be guided by insights specific to the data domain, numerical experimentation, or optimized through cross-validation to enhance the accuracy of predictive tasks based on $S$ values.

Our approach further incorporates the number of random walks per node $n$ and the length of these walks $w$ as parameters. Together with the dataset size $N$ and the embedding dimension $d$, these parameters specify the total number of permutation symbols \mbox{$M=N \times n \times (w-d+1)$} extracted from data. It is thus essential to have $M\gg d!$ to obtain a reliable estimate of the probability of all $d!$ possible ordinal patterns. Naturally, $w$ must exceed $d$ to accommodate the partitions $u_q$ within the sampled trajectories. Nonetheless, the choice between a large number of relatively short walks per node versus a small number of long walks per node tends to have minimal impact on estimating the $k$-nn permutation entropy, since the random walkers only remember their immediate previous position. Additionally, each replication of the random walks generates distinct sampled trajectories and permutation symbols, resulting in different estimates for the ordinal distributions and, consequently, different values for the $k$-nn permutation entropy $S$. Therefore, beyond ensuring $M$ significantly exceeds $d!$, the random walks must yield a representative sample of possible ordinal patterns extractable from the graph, thereby rendering variations in $S$ negligible for the analysis. The total number of walks of length $d$ can be calculated using powers of the graph adjacency matrix and may serve as a guide to set the values of $n$ and $w$. An alternative strategy is to incrementally increase their values until a desirable level of stability in the $k$-nn permutation entropy. In our applications with $N$ ranging from approximately $10^3$ to $10^5$ and $d$ between $3$ and $7$, we find that $n=w=10$ yields quite stable $S$ values. 

The last two parameters introduced by our approach are $\lambda$ and $\beta$, which influence the bias of the walkers. We set these parameters at fixed values ($\lambda=10$ and $\beta=0.001$) to steer the random walks toward visiting nodes progressively farther from the source node (depth-first sampling). This strategy proved to be more effective for sampling ordinal patterns across all our applications. However, adjusting $\lambda$ and $\beta$ might become relevant in scenarios characterized by a notable imbalance in the density of points or when the aim is a more detailed investigation of local data patterns. In such cases, the bias can be modified to favor nodes close to the source node (breadth-first sampling) by increasing $\beta$ and reducing $\lambda$. Alongside our article, we also introduce \href{github.com/hvribeiro/knnpe}{\texttt{knnpe}} --- an open-source Python module hosted at \url{github.com/hvribeiro/knnpe} designed to efficiently calculate the $k$-nn permutation entropy and allow comprehensive experimentation with its parameters. This tool can be installed from the Python Package Index (PyPI) via the command: \texttt{pip install knnpe}. 

In the subsequent sections, we demonstrate the application of the $k$-nn permutation entropy across a range of data structures, benchmarking its performance and noise robustness against established techniques in the data domain.

\section{Results}

\subsection{Unstructured data}

\begin{figure*}[ht!]
  \centering
  \includegraphics[width=1\textwidth, keepaspectratio]{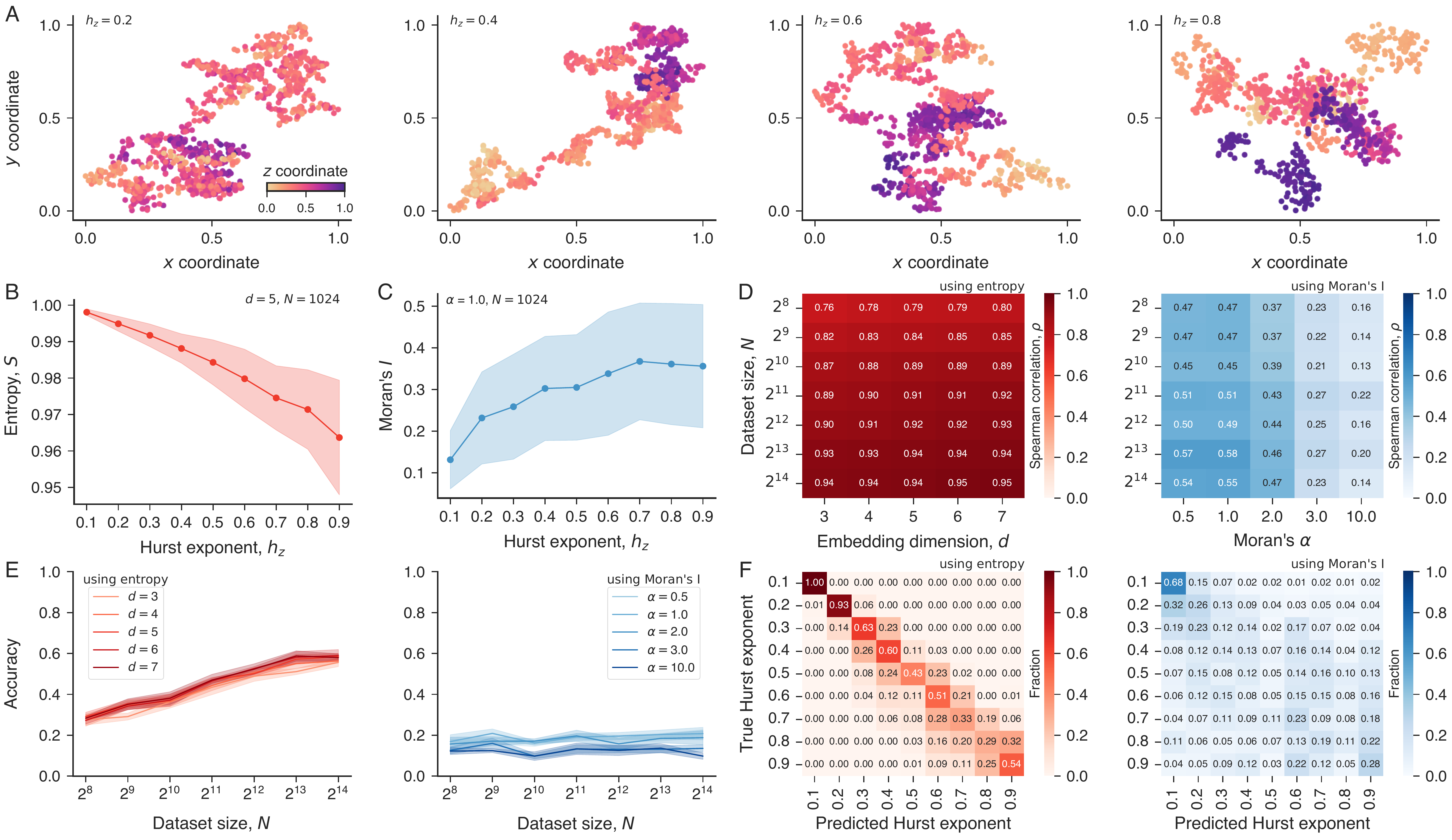}
  \caption{Detecting persistence patterns in irregularly distributed data via $k$-nearest neighbor permutation entropy. (A) Illustrations of irregularly distributed datasets where the $x$ and $y$ coordinates follow a standard Brownian motion ($h_x=h_y=0.5$), and the $z$ coordinate arises from a fractional Brownian motion characterized by different Hurst exponents $h_z$. Notably, lower values of $h_z$ yield patterns in which adjacent data points exhibit seemingly random $z_i$ values, whereas higher $h_z$ values lead to patterns in which adjacent points show local trends of increasing or decreasing $z_i$ values. (B) Relationship between $k$-nearest neighbor permutation entropy $S$ (with $d=5$) and Hurst exponent $h_z$. (C) Dependency of Moran's $I$ spatial autocorrelation (with $\alpha=1$) on Hurst exponent $h_z$. In the two previous panels, markers represent average values calculated from one hundred independent replicas of the process that generates the datasets (featuring $N=1024$ data points) for each $h_z\in\{0.1,0.2,\dots,0.9\}$, while shaded regions stand for one standard deviation confidence intervals. (D) Spearman correlation between the values of $S$ and $h_z$ (red shades) across various dataset sizes ($N$, rows) and embedding dimensions ($d$, columns), as well as the correlation between the values of $I$ and $h_z$ (blue shades) for different dataset sizes ($N$, rows) and distance exponents ($\alpha$, columns). (E) Accuracy of classification tasks aimed at predicting $h_z$ using $k$-nearest neighbor permutation entropy $S$ (red shades indicate different embedding dimensions $d$) and Moran's $I$ (blue shades indicate different values of $\alpha$) as a function of dataset size $N$. Shaded areas represent the standard deviation from the average accuracy levels estimated from ten independent realizations of the training process. (F) Examples of confusion matrices resulting from applying the learning algorithm to predict the Hurst exponent $h_z$ using entropy values $S$ (red shades, $N=2^{14}$ and $d=5$) and Moran's $I$ (blue shades, $N=2^{14}$ and $\alpha=1$).
  }
  \label{fig:2}
\end{figure*}

Our investigation begins by analyzing spatial datasets [$\vec{r}_i=(x_i,y_i)$ and $z_i$] constructed as 
\begin{equation}\label{eq:3dfbm}
    \begin{split}
        x_{i} &= x_{i-1} + \xi_{h_x} \\
        y_{i} &= y_{i-1} + \xi_{h_y} \\
        z_{i} &= z_{i-1} + \xi_{h_z} \\
    \end{split}\,,
\end{equation}
where $\xi_{h_x}$, $\xi_{h_y}$ and $\xi_{h_z}$ represent fractional Gaussian noises~\cite{mandelbrot1982fractal, mandelbrot1968fractional} with zero mean, unit variance, and Hurst exponents $h_x$, $h_y$ and $h_z$, respectively. These noise terms are numerically simulated using the Hosking method~\cite{hosking1984modeling}. Thus, the variables $x_i$, $y_i$, and $z_i$ denote fractional Brownian motions, with $x_i$ and $y_i$ serving as data coordinates and $z_i$ as the associated value. The Hurst exponents modulate the roughness of these stochastic processes. Exponents smaller than $1/2$ result in a time series exhibiting more frequent alternation in increment signs than expected by chance (anti-persistent behavior). Conversely, exponents greater than $1/2$ lead to time series in which increments maintain their signs more often than by chance (persistent behavior). When the Hurst exponent is $1/2$, these series correspond to conventional Brownian motions and larger exponents yield smoother time series.

We consider the case where the Hurst exponents for spatial coordinates are set to $h_x=h_y=1/2$, while the $h_z$ exponent varies from $0.1$ to $0.9$ with increments of size $0.1$. These choices maintain a constant spatial structure in the distribution of points and allow us to investigate whether the $k$-nn permutation entropy can identify variations in regularity in the $z_i$ values. Figure~\ref{fig:2}A shows four examples of simulated data for different $h_z$ values. Lower values of $h_z$ generate more random patterns, whereas higher values yield spatial patterns characterized by similar values and spatial trends among adjacent data points. We create an ensemble comprising one hundred independent replications of these datasets for each $h_z$ and different dataset sizes $N=2^8,2^9,\dots,2^{14}$. For each dataset size, we assess how the $k$-nn permutation entropy $S$ depends on the Hurst exponent $h_z$ for different embedding dimensions $d$, as illustrated in Figure~\ref{fig:2}B for $N=1024$ and $d=5$. In this figure, markers correspond to average values of $S$, and the shaded areas represent the confidence interval of one standard deviation. The $k$-nn permutation entropy monotonically decreases with the Hurst exponent $h_z$ and can thus distinguish among different degrees of spatial structure in data. Higher entropy values correspond to more randomness, while lower values reflect persistent spatial patterns associated with higher Hurst exponents. 

We compare these findings with Moran's $I$~\cite{moran1950notes}, which is argued to be the most widely used measure of spatial autocorrelation~\cite{getis2008history}. This measure is essentially a spatially-weighted extension of the Pearson correlation coefficient and is defined as
\begin{equation}
    I = \frac{N}{W_0} \frac{\sum_i^N\sum_j^N \omega_{ij} (z_i-\bar{z})(z_j-\bar{z})}{\sum_i^N(z_i-\bar{z})^2 }\,,
\end{equation}
where $\bar{z}$ denotes the average of $z_i$, $\omega_{ij}$ is a weight between data points $z_i$ and $z_j$, and $W_0=\sum_i^N\sum_j^N \omega_{ij}$ is a normalization constant. Moran's $I$ usually ranges from $-1$ to $1$, with $I\to I_0={-1}/{(N-1)}$ for uncorrelated spatial data, while $I>I_0$ and $I<I_0$ respectively indicate positive and negative spatial correlations~\cite{de1984extreme}. The simplest form of the weight matrix sets $\omega_{ij}=1$ for immediate neighbors and $\omega_{ij}=0$ otherwise (with $\omega_{ii}=0$). Another prevalent approach involves defining the weight $\omega_{ij}$ by a distance decay function $\omega_{ij}=1/d_{ij}^\alpha$, where $d_{ij}=\sqrt{(x_i-x_j)^2+(y_i-y_j)^2}$ is the distance between the points $z_i$ and $z_j$, and $\alpha$ represents an exponent~\cite{gittleman1990adaptation}. We adopt this latter approach and use different values of $\alpha$ to examine how $I$ depends on the Hurst exponent $h_z$. As illustrated in Figure~\ref{fig:2}C for $\alpha=1$ and $N=1024$, $I$ values tend to rise with $h_z$. However, unlike the $k$-nn permutation entropy $S$, the relationship is not monotonous, with $I$ values reaching a plateau around $h_z=0.7$. Moreover, $I$ displays considerably greater relative dispersion than $S$, as evidenced by the extensive shaded area representing the confidence interval of one standard deviation. These observations indicate that Moran's $I$ is less effective in identifying varying degrees of spatial structure in our datasets compared to the $k$-nn permutation entropy $S$. 

To systematically contrast both measures, we calculate the Spearman correlation between $S$ and $h_z$, organizing the results by dataset size $N$ and embedding dimension $d$. We then compare these findings with the Spearman correlation between $I$ and $h_z$, categorized by dataset size $N$ and distance exponent $\alpha$. As depicted in Figure~\ref{fig:2}D, the correlations between $S$ and $h_z$ are significantly stronger than those between $I$ and $h_z$, irrespective of dataset size, embedding dimension, or distance exponent. Additionally, we evaluate the predictive efficacy of $S$ and $I$ in machine learning tasks employing the nearest neighbors classifier~\cite{james2013introduction} to predict $h_z$ values using each measure as a predictive feature. We split the data into test (20\%) and training sets, stratifying by $h_z$ values, train the classifiers using a three-fold cross-validation approach to fine-tune the number of neighbors in the learning algorithm, and then determine the accuracy of the predictions on the test set. We further repeat the train-test split and the training process across ten independent instances, calculating the average and the standard deviation of the accuracy on the test sets. Figure~\ref{fig:2}E shows the average accuracy as a function of the dataset size obtained using the entropy values for different embedding dimensions $d$, compared with the corresponding values calculated with Moran's $I$ for various $\alpha$ values. These results indicate that the $k$-nn permutation entropy significantly enhances the accuracy of predicting the Hurst exponent $h_z$ compared to Moran's $I$, which generally falls below the 20\% accuracy threshold. We also compute the confusion matrix for these classification talks on the test sets, as depicted in Figure~\ref{fig:2}F. The diagonal band in the matrix indicates that incorrect classifications made by the algorithm trained with $S$ values typically yield Hurst exponents close to the actual value. Conversely, the confusion matrix from the algorithm trained with $I$ values lacks a similar diagonal pattern, underscoring the inferior predictive capability of $I$ relative to the $k$-nn permutation entropy.

\begin{figure*}[ht!]
  \centering
  \includegraphics[width=1\textwidth, keepaspectratio]{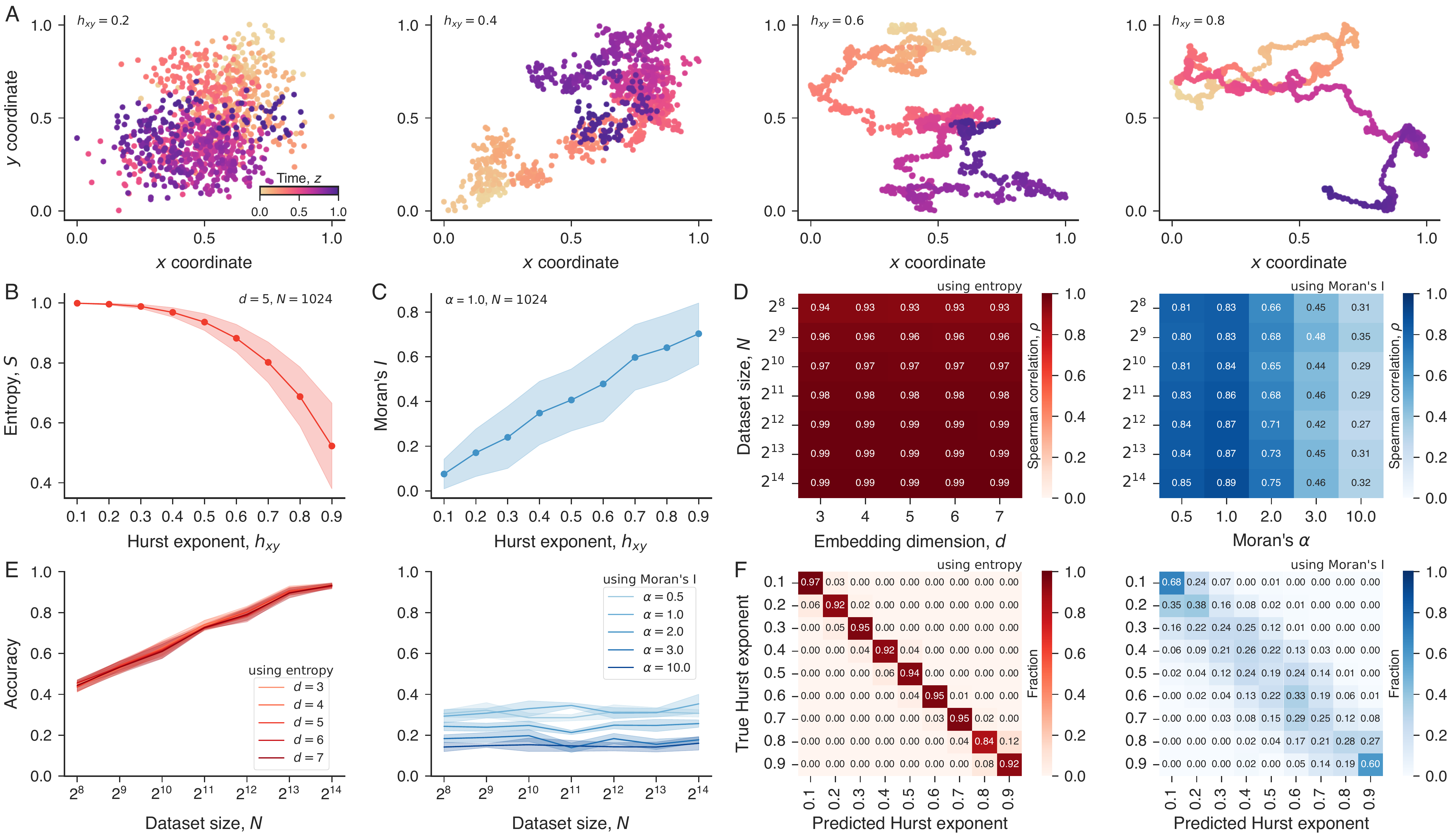}
  \caption{Detecting spatial patterns using $k$-nearest neighbor permutation entropy. (A) Illustrations of irregularly distributed datasets where the $x$ and $y$ coordinates correspond to two-dimensional fractional Brownian motions with different Hurst exponents $h_{xy}$, and the $z$ coordinate denotes time $t$ (a sequential index for each step in the two-dimensional motion). Values of $h_{xy}<0.5$ produce an anti-persistent behavior characterized by frequent sign alternation in position increments along both axes. This behavior disrupts any clear sequence in the times $t$ of adjacent points, owing to prevalent back-and-forth movement. In contrast, position increments often show the same sign for $h_{xy}>0.5$, resulting in persistent motion and a tendency for times $t$ of neighboring points to follow a sequential order. (B) Relationship between $k$-nearest neighbor permutation entropy $S$ (with $d=5$) and Hurst exponent $h_{xy}$ of two-dimensional fractional Brownian motions. (C) Dependency of Moran's $I$ spatial autocorrelation (with $\alpha=1$) on Hurst exponent $h_{xy}$. In the two previous panels, markers represent average values calculated from one hundred independent replicas of the process that generates the datasets (featuring $N=1024$ data points) for each $h_{xy}\in\{0.1,0.2,\dots,0.9\}$, while shaded areas stand for standard deviation confidence intervals. (D) Spearman correlation between the values of $S$ and $h_{xy}$ (red shades) across various dataset sizes ($N$, rows) and embedding dimensions ($d$, columns), as well as the correlation between the values of $I$ and $h_{xy}$ (blue shades) for different dataset sizes ($N$, rows) and distance exponents ($\alpha$, columns). (E) Accuracy of classification tasks aimed at predicting $h_{xy}$ values using $k$-nearest neighbor permutation entropy $S$ (red shades for different $d$) and Moran's $I$ (blue shades for different $\alpha$) as a function of dataset size $N$. Shaded areas represent the standard deviation of accuracy levels obtained from ten independent realizations of the training process. (F) Examples of confusion matrices resulting from applying the learning algorithm to predict the Hurst exponent $h_{xy}$ of two-dimensional fractional Brownian motions using entropy values $S$ (red shades, $N=2^{14}$ and $d=5$) and Moran's $I$ (blue shades, $N=2^{14}$ and $\alpha=1$).
  }
  \label{fig:3}
\end{figure*}

In our second application involving unstructured data, we explore whether the $k$-nn permutation entropy can identify structural changes in the spatial distribution of data points when the pattern associated with the $z_i$ values remains constant. To investigate this, we adapt the model presented in Equation~\ref{eq:3dfbm} to create spatial datasets [$\vec{r}_i=(x_i,y_i)$ and $z_i$] as follows:
\begin{equation}\label{eq:2dfbm}
    \begin{split}
        x_{i} &= x_{i-1} + \xi_{h_{xy}}^{(x)} \\
        y_{i} &= y_{i-1} + \xi_{h_{xy}}^{(y)} \\
        z_{i} &= i \\
    \end{split}\,,
\end{equation}
where $\xi_{h_{xy}}^{(x)}$ and $\xi_{h_{xy}}^{(y)}$ are fractional Gaussian noises~\cite{mandelbrot1982fractal, mandelbrot1968fractional} with zero mean, unit variance, and an identical Hurst exponent $h_{xy}$. The datasets generated via this model emulate a two-dimensional fractional Brownian motion in the data coordinates [$\vec{r}_i=(x_i,y_i)$], while the $z_i$ values simply represent the time step indices. As illustrated in Figure~\ref{fig:3}A, lower $h_{xy}$ values result in antipersistent behavior in the data coordinates, leading to more irregular patterns associated with the $z_i$ values. In contrast, higher $h_{xy}$ values induce a persistent behavior in the data coordinates, yielding more regular patterns characterized by adjacent $z_i$ values that display increasing or decreasing trends. 

We consider $h_{xy}=0.1,0.2,\dots,0.9$ to generate an ensemble comprising one hundred independent replications of these datasets for each $h_{xy}$ and dataset sizes $N=2^8,2^9,\dots,2^{14}$. Using these datasets, we estimate the $k$-nn permutation entropy $S$ as a function of the Hurst exponent $h_{xy}$ for various embedding dimensions $d$, in addition to analyzing how the Moran's $I$ depends on $h_{xy}$ for distinct values of the distance exponent $\alpha$. Figures~\ref{fig:3}B and~\ref{fig:3}C show these relationships for $N=1024$, with entropy values calculated using $d=5$ and $I$ values determined with $\alpha=1$. Consistent with our expectations, we note that $S$ diminishes with increasing $h_{xy}$ while $I$ rises. Moran's $I$ exhibits a roughly linear association, whereas the $k$-nn permutation entropy $S$ displays a non-linear relationship marked by a rapid decrease for $h_{xy}>0.5$. Notably, the relative dispersion, quantified by one standard deviation confidence intervals, is considerably smaller for $S$ than for $I$, especially for lower Hurst exponents. We assess the quality of these relationships by computing the Spearman correlation, categorizing the results by $N$ and also by $d$ (for entropy) and $\alpha$ (for Moran's $I$). Figure~\ref{fig:3}D presents these correlations, indicating that the association between $S$ and $h_{xy}$ is substantially more stable than that between $I$ and $h_{xy}$, regardless of the dataset size or the parameters $d$ and $\alpha$. Applying the same machine learning approach used in our first application, we compare the efficacy of $S$ and $I$ in predicting the Hurst exponents $h_{xy}$. Figure~\ref{fig:3}E displays the average accuracy of these predictive tasks as a function of the dataset size, using entropy values for distinct embedding dimensions and Moran's $I$ calculated for various distance exponents. Once again, the $k$-nn permutation entropy performance notably surpasses that of Moran's $I$, irrespective of the dataset size or the parameters $d$ and $\alpha$. This superiority is further supported by the confusion matrices in Figure~\ref{fig:3}F, where incorrect predictions made by the algorithm trained with entropy values are typically one step removed from the true Hurst exponent, resulting in an almost perfect diagonal pattern. Conversely, the algorithm trained with Moran's $I$ presents a considerably wider diagonal band, reflecting the lesser efficacy of this metric for the task.

\subsection{Irregularly sampled and noisy time series}

\begin{figure*}[ht!]
  \centering
  \includegraphics[width=1\textwidth, keepaspectratio]{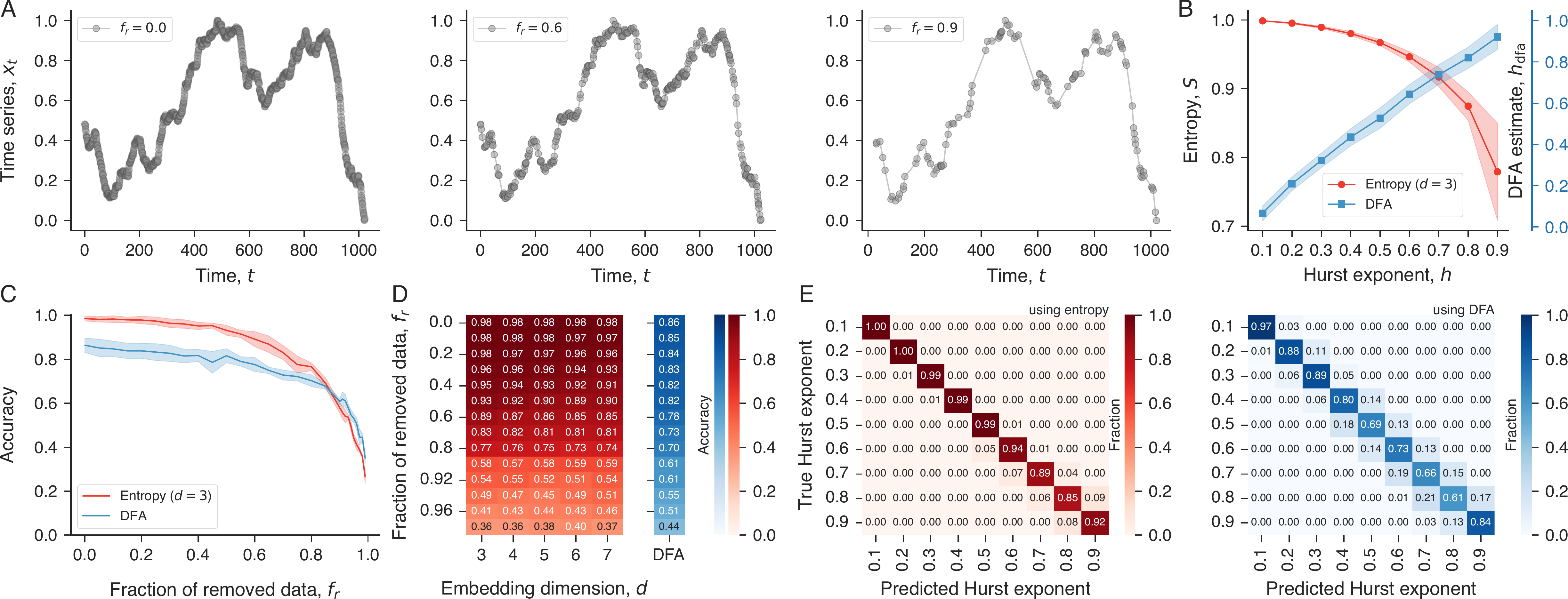}
  \caption{Applying $k$-nearest neighbor permutation entropy to irregularly sampled time series. (A) Examples of fractional Brownian motion time series illustrating the impact of randomly removing different fractions $f_r$ of data points ($h=0.8$, $N=1024$, and $f_r\in\{0,0.6,0.9\}$ in these instances). (B) Relationship between $k$-nearest neighbor permutation entropy $S$ (with $d=3$) and Hurst exponent $h$ (red circles), along with the association between estimated Hurst exponent from DFA $h_{\text{dfa}}$ and the actual Hurst exponent $h$ (blue squares). In both curves, markers indicate average values estimated from one hundred distinct realizations of fractional Brownian motions, each with a length $N=2^{13}=8192$ and a fraction of removed points $f_r=0.45$. Shaded areas represent one standard deviation confidence intervals. (C) Analysis of the accuracy in classification tasks aimed at predicting the Hurst exponents ($h\in\{0.1,0.2,\dots,0.9\}$) using $k$-nearest neighbor permutation entropy $S$ (red curve) versus estimates from DFA ($h_{\text{dfa}}$, blue curve) as a function of the fraction of removed data points $f_r$. Shaded regions represent the standard deviation of accuracy levels based on ten independent realizations of the training process. (D) Accuracy in predicting the Hurst exponent $h$ for various fractions of removed data points $f_r$ (rows) using entropy $S$ with different embedding dimensions $d$ (columns in red shades) and using the DFA (column in blue shades). (E) Example of confusion matrices resulting from applying the learning algorithm to predict the value of $h$ using the entropy $S$ for $d=3$ (red shades) and DFA estimates $h_{\text{dfa}}$ (blue shades) with 45\% of data points randomly excluded from time series.
  }
  \label{fig:4}
\end{figure*}

The $k$-nn permutation entropy further provides a natural method for incorporating amplitude information and time gaps when analyzing one-dimensional time series $\{x_t\}$. To achieve this, we consider the spatial coordinates as $\vec{r}_t = (x_t, t)$ and assign time series values $x_t$ to the $z_t$ values. This approach ensures that amplitude and time gaps are accounted for in the $k$-nearest neighbor graph construction. We evaluate the efficacy of this method in two applications. In the first, we analyze time series derived from fractional Brownian motions with a Hurst exponent $h$, where a specific fraction $f_r$ of data points is randomly removed, resulting in an irregularly sampled series, as depicted in Figure~\ref{fig:4}A for various $f_r$ and $h=0.8$. We generate an ensemble with one hundred time series, each initially containing $N=2^{13}$ points and Hurst exponents ranging from $h=0.1$ to $h=0.9$ in steps of $0.1$. We then explore the relation between the $k$-nn permutation entropy $S$ and the Hurst exponent $h$ for different fractions of removed points ($f_r=0, 0.1, \dots, 0.9, 0.92, \dots, 0.98$) and several embedding dimensions $d$. Figure~\ref{fig:4}B illustrates this relationship for $f_r=0.45$ and $d=3$ (red circles, left $y$ axis). Additionally, we apply the detrended fluctuation analysis (DFA)~\cite{peng1994mosaic} to our ensemble of irregularly sampled time series, one of the most widely used and reliable methods for estimating Hurst exponents~\cite{shao2012comparing}. The DFA involves calculating the root-mean-square fluctuation function $F(m)$ over non-overlapping time series partitions of length $m$ after removing a polynomial trend (in our case, we use linear polynomials). For self-similar time series, such as those from fractional Brownian motions, $F(m)$ presents a power-law dependence on the partition length $m$, expressed as $F(m)\sim m^{h_{\text{dfa}}}$, where $h_{\text{dfa}}$ is the estimated Hurst exponent. We determine $h_{\text{dfa}}$ by computing the slope of the linearized version of this power-law relationship ($\log F(m) \sim h_{\text{dfa}} \log m$) using the ordinary least squares method. Moreover, the DFA does not consider the time gaps of our irregularly sampled time series; instead, the time value is replaced by a sequential number denoting the observation index. Figure~\ref{fig:4}B also illustrates the relation between $h_{\text{dfa}}$ and the actual Hust exponent $h$ when 45\% of the data points are removed (blue circles, right $y$ axis). The association between $h_{\text{dfa}}$ and $h$ is approximately linear, while the values of $S$ decreases non-linearly with $h$, similar to the behavior observed in two-dimensional fractional Brownian motion (Figure~\ref{fig:3}B). It is worth noting that the variability in $S$ values (represented by the one standard deviation confidence intervals in Figure~\ref{fig:4}B) is significantly smaller than that observed for $h_{\text{dfa}}$, especially for lower Hurst exponents.

This result suggests that the $k$-nn permutation entropy potentially offers greater precision in predicting the Hurst exponents compared to DFA. To rigorously evaluate this hypothesis, we employ the same machine learning methodology of our previous investigations for classifying Hurst exponents $h$ using either the entropy values $S$ (with $d=3$) or the DFA-derived estimates $h_{\text{dfa}}$, across varying fractions of data removal $f_r$. Figure~\ref{fig:4}C shows the average accuracy for each method as a function of $f_r$. Although both classifiers achieve high accuracy levels, the classifier based on entropy values outperforms the DFA-based classifier, particularly when $f_r<0.5$, where the entropy-based classifier achieves an average accuracy exceeding 95\%. Accuracy sharply declines for both classifiers when approximately 80\% or more data points are omitted from the series, with the DFA method showing slightly better precision than the $k$-nn permutation entropy under these conditions. We also verify that changes in the embedding dimension $d$ do not significantly affect the classification accuracy, as shown in Figure~\ref{fig:4}D. Additionally, we compare the confusion matrices from classifiers trained with both metrics, as illustrated in Figure~\ref{fig:4}E for $f_r=0.45$ and $d=3$. In general, these matrices exhibit a pronounced diagonal pattern that indicates the classifiers' effectiveness. However, the diagonal elements of the confusion matrices using $S$ values are closer to one, highlighting superior classifier performance, especially at higher Hurst exponents where classifiers using $h_{\text{dfa}}$ display increased confusion.

\begin{figure*}[ht!]
  \centering
  \includegraphics[width=1\textwidth, keepaspectratio]{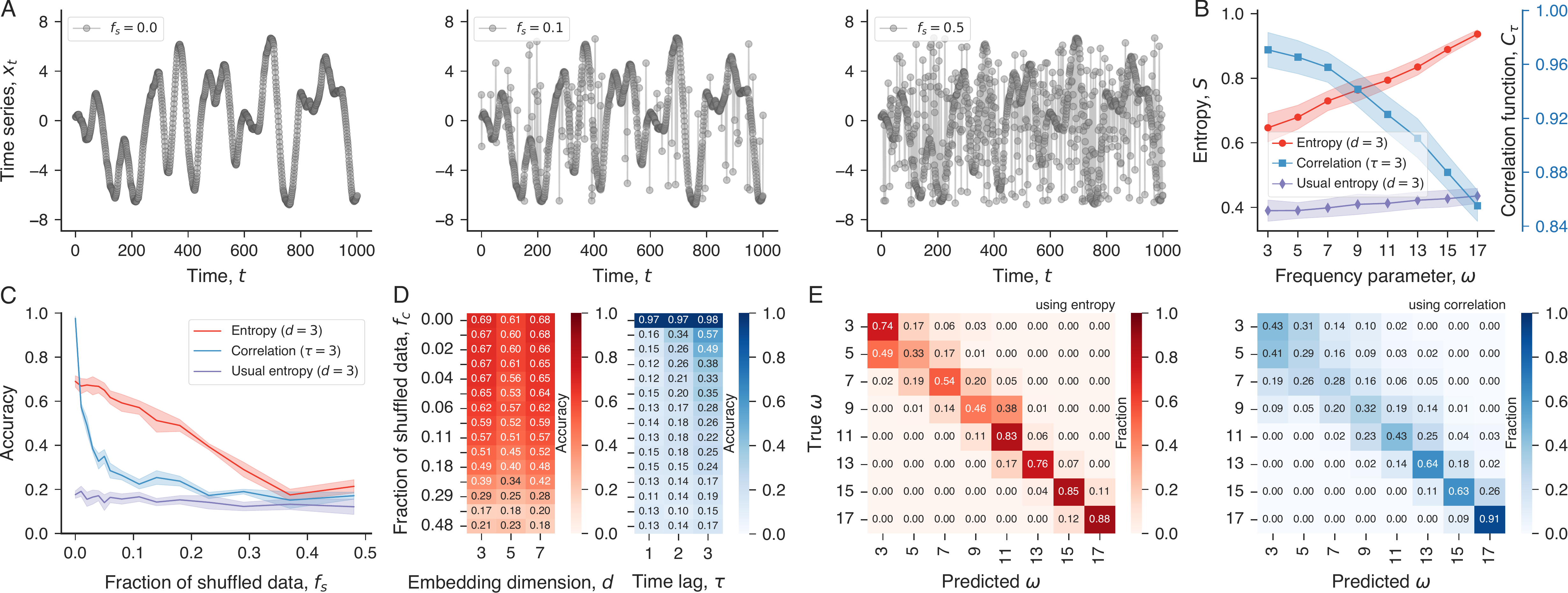}
  \caption{Investigating harmonic noise time series using $k$-nearest neighbor permutation entropy. (A) Illustrations of harmonic noises time series demonstrating the effect of swapping positions of randomly selected pairs of points at different fractions $f_s\in\{0,0.1,0.5\}$ ($\omega=5$ and $\Gamma=3$ in these instances). (B) Relationship between the $k$-nearest neighbor permutation entropy $S$ (with $d=3$) and harmonic noise frequency $\omega$ (red circles), juxtaposed with the dependency of the correlation function $C_\tau$ (with lag $\tau=3$, blue squares) and the usual permutation entropy (with $d=3$, purple diamonds) on the $\omega$ parameter (both for $f_s=0.01$). (C) Comparative analysis of the accuracy in predicting the frequency parameter $\omega$ using $k$-nearest neighbor permutation entropy (red curve, $d=3$), correlation function $C_\tau$ (blue curve, $\tau=3$), and usual permutation entropy (purple curve, $d=3$) as a function of the fraction of shuffled data $f_s$. Shaded regions represent one standard deviation in accuracy estimations calculated from ten independent realizations of the training process. (D) Evaluation of the accuracy in predicting the harmonic noise frequency ($\omega\in\{3,5,\dots,17\}$) across varying $f_s$ values (rows) using $k$-nearest neighbor permutation entropy with different embedding dimensions (columns, red shades) and the correlation function with different time lags (columns, blue shades). (E) Examples of confusion matrices resulting from applying the learning algorithm to predict the parameter $\omega$ using $k$-nearest neighbor permutation entropy with $d=3$ (red shades) and correlation function with $\tau=3$ (blue shades) in time series with $1\%$ of shuffled data.
  }
  \label{fig:5}
\end{figure*}

In our second application, we examine time series derived from the harmonic noise, which corresponds to a harmonic oscillator driven by a Gaussian noise, expressed as the Langevin equations~\cite{geier1990harmonic}
\begin{equation}\label{eq:langevin_harm}
\frac{dx}{dt^\prime} = v~~\text{and}~~\frac{dv}{dt^\prime} = - \Gamma v - \Omega^2 x + \sqrt{2 \varepsilon}\, \Omega^2 \xi(t^\prime)\,,
\end{equation}
where $\Gamma$, $\Omega$ and $\epsilon$ are model parameters, while $\xi(t^\prime)$ stands for a Gaussian white noise. The autocorrelation function $C(\tau)=\langle x(t^\prime) x(t^\prime+\tau)\rangle$ exhibits an oscillatory decay~\cite{geier1990harmonic}
\begin{equation}\label{eq:harm_cor}
C(\tau) = \frac{\varepsilon \Omega^2}{\Gamma} \exp\left(-\frac{\Gamma}{2}\tau\right) \left[\cos(\omega \tau) + \frac{\Gamma}{2\omega}\sin(\omega \tau)\right]\,,
\end{equation}
where $\omega = \sqrt{\Omega^2 - (\Gamma/2)^2}$ is the oscillation frequency. This harmonic noise merges periodic and stochastic behaviors, serving as a comprehensive model for evaluating the discriminative power of the $k$-nearest neighbor permutation entropy on time series. We numerically integrate the Langevin equations using the Euler method with a step size $dt^\prime=10^{-2}$ to closely match the exact autocorrelation function and its numerical estimates. The integration proceeds until a maximum time $t^\prime_{\max}=10$, yielding time series $\{x_t\}_{t=1,\dots,N}$ with $N=1000$ elements. We generate an ensemble comprising one hundred independent replicas of these time series with $\epsilon=1$ and $\Gamma=3$, for each $\omega$ varying between $3$ and $17$ in increments of $2$. These parameters produce an autocorrelation function resembling the motion of an underdamped harmonic oscillator and are chosen because the usual permutation entropy is known to struggle in identifying changes in the oscillation frequency within this range~\cite{ribeiro2017characterizing}. In addition, to evaluate robustness to noise of the $k$-nn permutation entropy, we randomly swap pairs of elements in these time series at various fractions $f_s$ of their length, ranging from $0$ to $0.48$ in 14 increasingly spaced samples. Figure~\ref{fig:5}A depicts examples of these time series for $\omega=5$ and three different fractions $f_s$.

We calculate the dependency of the $k$-nearest neighbor permutation entropy on the oscillation frequency $w$ for each fraction $f_s$, comparing the results with the relationships derived from the autocorrelation function $C_\tau$ (numerically estimated from time series) for different time lags $\tau$ and the usual permutation entropy. These curves, depicted in Figure~\ref{fig:5}B for $f_s=0.01$ and $d=\tau=3$, demonstrate that both the $k$-nearest neighbor permutation entropy and the usual permutation exhibit an increasing trend with $\omega$ values, whereas the autocorrelation diminishes as $\omega$ increases. However, the relative rate of change in the usual permutation entropy is considerably smaller than those observed for the other two metrics. The usual permutation entropy values as a function of $\omega$ are almost entirely within one standard deviation confidence intervals, underscoring the challenge of differentiating harmonic noises with this metric~\cite{ribeiro2017characterizing}. When comparing the relationships derived from the $k$-nearest neighbor permutation entropy and the autocorrelation, we note that the relative dispersion in $S$ values is substantially lower than in $C_\tau$ values, especially at high-frequency values. 

To methodically compare the three metrics, we once again apply our machine learning approach, using each quantifier to classify the values of $\omega$ at different fractions of shuffled data $f_s$. Figure~\ref{fig:5}C shows the average accuracy of each classifier as a function of $f_s$ for $d=\tau=3$. The usual permutation entropy exhibits the worst performance, with accuracy levels consistently below $0.2$, a result that does not improve with larger embedding dimensions. Without external noise in the time series ($f_s=0$), the classifiers trained with the autocorrelation function $C_\tau$ achieve near-perfect accuracy (97\%) and significantly outperform those trained with the $k$-nearest neighbor permutation entropy (69\%). However, the accuracy of $C_\tau$-based classifiers is highly susceptible to external noise, falling below that of the $k$-nearest neighbor permutation entropy when only 1\% of the time series points are shuffled. The accuracy of $C_\tau$-based classifiers further declines to levels comparable to those of the usual permutation entropy when $f_s=0.1$, while the $k$-nearest neighbor permutation entropy remains robust against the addition of external noise. These results are not significantly influenced by variations in the embedding dimension $d$ or the time lag $\tau$, as illustrated in Figure~\ref{fig:5}D. With minimal external noise, the confusion matrices of classifiers trained with $S$ or $C_\tau$ are similar to the ones presented in Figure~\ref{fig:5}E for $f_s=0.01$, indicating that the $k$-nearest neighbor permutation entropy significantly improves classifications at lower frequency values. The enhanced noise resilience of the $k$-nearest neighbor permutation entropy is attributable to the fact that ordinal patterns are not strictly derived from sequential elements in time series. Instead, they are calculated from time series samples obtained via random walks over a $k$-nearest neighbor graph, which connects nearby points within the time series space, capturing thus the dominant time series patterns even under high levels of external noise. Points deviating significantly from these dominant patterns are sampled less frequently by the walker trajectory, thereby enhancing the noise robustness of our approach.

\subsection{Liquid crystal textures}

\begin{figure*}[ht!]
  \centering
  \includegraphics[width=0.9\textwidth, keepaspectratio]{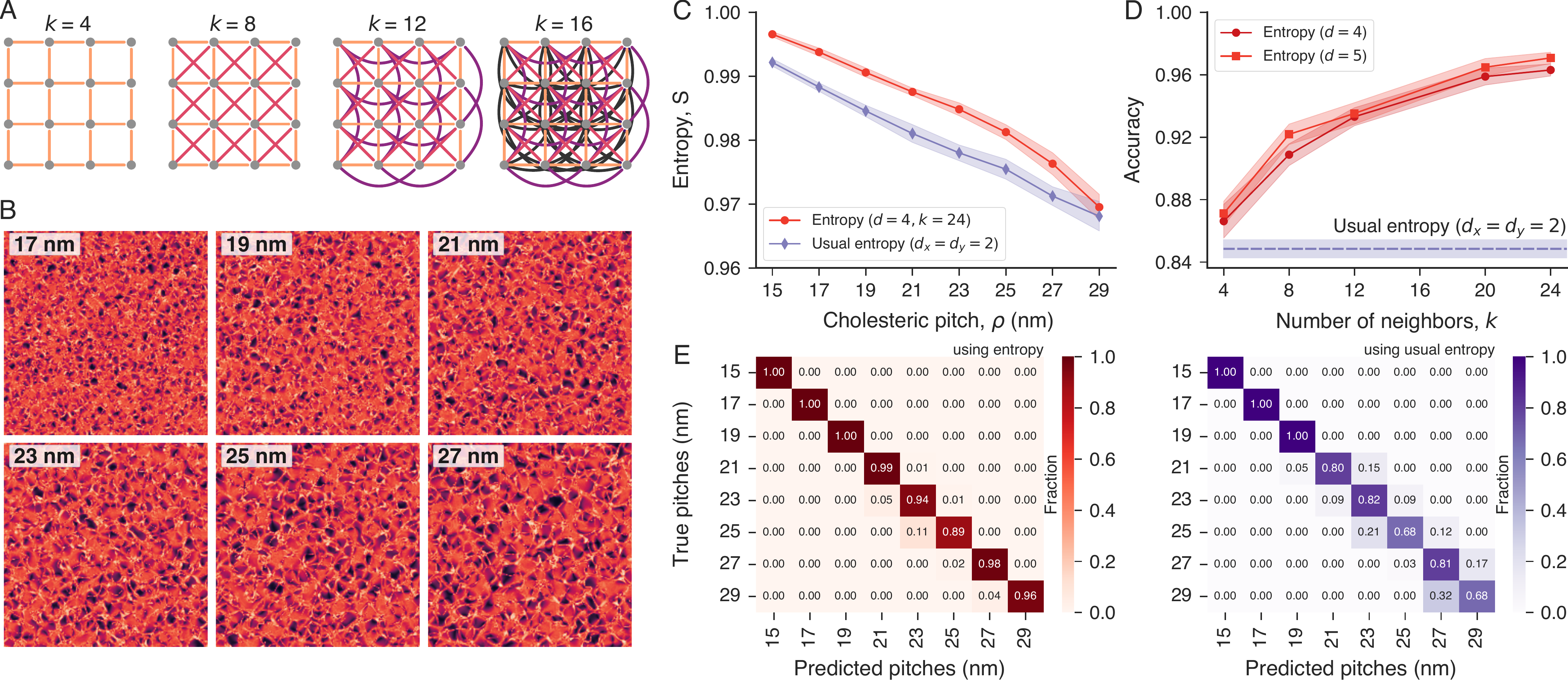}
  \caption{Using $k$-nearest neighbor permutation entropy for image analysis. (A) Illustrations of regular graphs formed by connecting pixels in a $4\times4$ image to their first ($k=4$), second ($k=8$), third ($k=12$), and fourth ($k=16$) neighbors. (B) Optical textures of cholesteric liquid crystals with different pitches $\rho$. (C) Relationship between $k$-nearest neighbor permutation entropy $S$ (with $d=4$ and $k=24$) and cholesteric pitch $\rho$ (red circles), juxtaposed with results using conventional two-dimensional permutation entropy (purple diamonds). Markers represent average values calculated from 355 image textures ($150\times150$ pixels) per pitch, with shaded regions indicating one standard deviation confidence bands. (D) Accuracy in predicting cholesteric pitch using $k$-nearest neighbor permutation entropy as a function of the number of neighbors $k$ for embedding dimensions $d=4$ (dark red circles) and $d=5$ (light red squares). The dashed horizontal line represents the accuracy attained using conventional two-dimensional permutation entropy ($d_x=d_y=2$). Shaded areas display one standard deviation of accuracy levels estimated from ten independent realizations of the training process. (E) Examples of confusion matrices obtained when applying the learning algorithm to predict cholesteric pitch using $k$-nearest neighbor permutation entropy ($d=5$ and $k=24$, red shades) versus usual two-dimensional permutation entropy with ($d_x=d_y=2$, blue shades).
  }
  \label{fig:6}
\end{figure*}

The improved performance and enhanced noise resilience achieved by applying the $k$-nn permutation entropy to regular time series motivate us to extend this method to image analysis. Traditionally, applying permutation entropy to images involves calculating ordinal patterns from flattened two-dimensional partitions of size $d_x \times d_y$ (the embedding dimensions) across regular arrays or matrices that represent pixel intensities of an image~\cite{ribeiro2012complexity}. The $k$-nn permutation entropy allows us to extract ordinal patterns from an image array $\{x_{ij}\}^{j=1,\dots,N_y}_{i=1,\dots,N_x}$, where $x_{ij}$ denotes the pixel intensity at the position $(i,j)$, in more versatile and potentially informative manners. This can be accomplished by considering the spatial coordinates as $\vec{r}_t = (i, j)$ and associating the pixel intensities $x_{ij}$ to the $z_t$ values, where $t=1,\dots,N$ enumerate all $N = N_x N_y$ pixels in an image. By varying the number of neighbors $k$ in our method, we create nearest neighbor graphs wherein each pixel is linked to a different number of neighboring pixels. For example, as depicted in Figure~\ref{fig:6}A, we can generate graphs in which the first ($k=4$), second ($k=8$), third ($k=12$), and fourth ($k=16$) nearest neighbors are connected, and so on. Consequently, sampling pixel values using random walkers on these graphs allows us to explore ordinal patterns beyond gridded partitions and from multiple spatial scales. 

Our investigation into the efficacy of this approach for image analysis focuses on examining textures of cholesteric liquid crystals with different pitch lengths that were previously studied using the usual permutation entropy~\cite{sigaki2019estimating}. Liquid crystals represent an intermediate state between the solid crystalline and liquid phases, where molecules have an average preferential direction defined by a vector known as the director~\cite{de1993physics}. Cholesteric liquid crystals are chiral materials that organize themselves by forming a helical structure, within which the director varies periodically, and the pitch length $\rho$ corresponds to the distance over which a full rotation is completed. For our analysis, we use 355 textures of size $150\times150$ for each pitch length ranging from $\rho=15$~nm to $\rho=29$~nm in steps of $2$~nm (for further details on these textures, we refer to Sigaki \textit{et al.}~\cite{sigaki2019estimating}). Figure~\ref{fig:6}B displays examples of these textures for six distinct pitch lengths, where each pixel represents the light intensity transmitted through the samples. 

We calculate the $k$-nn permutation entropy $S$ for all textures in our dataset, using the number of neighbors $k=4,8,\dots,24$ (corresponding to first through sixth neighbors) and embedding dimensions $d=4,5$. Subsequently, we evaluate the relationship between the average values of $S$ and the pitch lengths $\rho$, comparing these results with those from the usual two-dimensional permutation entropy for $d_x=d_y=2$. As shown in Figure~\ref{fig:6}C for $k=24$ and $d=4$, both entropy measures exhibit a similar decreasing trend with pitch length. The shaded areas represent the standard deviation of each quantifier across the textures for each pitch. A detailed examination of these curves reveals that variability in the $k$-nn permutation entropy is marginally lower than in the usual permutation entropy. To verify whether this slight difference enhances classification efficacy for the $k$-nn permutation entropy, we apply the same machine learning methodology of our previous investigations to categorize cholesteric textures, using each entropy measure as a predictive feature. Results of Figure~\ref{fig:6}D show the average accuracy of classifiers trained with $S$ values and an increasing number of neighbors, compared to the average accuracy achieved with the usual permutation entropy. Using the $k$-nn permutation entropy yielded a marginal improvement in accuracy for $k=4$ (86\% versus 85\% for the conventional entropy). Nonetheless, as the number of neighbors used for computing the $k$-nn permutation entropy increases, the disparity between the two methods becomes pronounced. Notably, the accuracy reaches 96\% for $k=24$, a performance comparable to a more sophisticated method based on deep convolutional neural networks~\cite{sigaki2020learning}. Figure~\ref{fig:6}E further compares typical confusion matrices from classifiers trained with both entropy measures, highlighting that the $k$-nn permutation entropy mainly improves the classification accuracy for high pitch values. Therefore, ordinal patterns originating from the sampling strategy introduced by the $k$-nn permutation entropy yield more information on the image structure, which translates into improved classification performance.

\section{Conclusion}

We have introduced an approach that generalizes the permutation entropy method to unstructured data. Our method is based on the construction of nearest neighbor graphs connecting closer data points and establishing neighborhood relations among them. Random walks over these graphs sample the values associated with each data point and allow the extraction of ordinal patterns and their distributions. The Shannon entropy of these ordinal distributions defines a quantifier we have designated as the $k$-nearest neighbor permutation entropy or $k$-nn permutation entropy. This new tool thus allows the characterization of data types beyond the traditional gridded data, such as time series or images, which are typically analyzed with the permutation entropy. We have tested the efficacy of this novel tool in examining patterns in irregularly spaced data derived from controlled \textit{in silico} experiments, demonstrating its effectiveness to detect changes in these patterns and its superior classification performance over a widely used measure of spatial autocorrelation known as the Moran's $I$.

Beyond expanding the applicability of ordinal methods to unstructured data, we have also demonstrated that the $k$-nn permutation entropy can be successfully applied to time series and images. Indeed, we have verified that nearest neighbor graphs used to calculate the $k$-nn permutation entropy inherently integrate information about amplitude and time gaps in the analysis of regular data. This inclusion has significantly enhanced noise resilience and predictive capacity of our method compared to the usual permutation entropy. Additionally, we have verified that the $k$-nn permutation entropy outperforms celebrated methods in complex systems, such as the detrended fluctuation analysis. 

Finally, we posit that $k$-nn permutation entropy has the potential for the analysis of multidimensional data types, including point cloud data, which is prevalent in LiDAR sensor applications and medical imaging, as well as in the characterization of point processes and geospatial data. Our approach further paves the way for the extension of a variety of other ordinal methods to unstructured data, addressing the pressing need for methodologies capable of efficiently extracting interpretable and valuable information from the increasingly voluminous and complex datasets used in academia and industry.

\begin{acknowledgments}
The authors acknowledge the support of the Coordena\c{c}\~ao de Aperfei\c{c}oamento de Pessoal de N\'ivel Superior (CAPES), the Conselho Nacional de Desenvolvimento Cient\'ifico e Tecnol\'ogico (CNPq -- Grant 303533/2021-8), the Consejo Nacional de Investigaciones Científicas y Técnicas (CONICET, Argentina), and the Slovenian Research Agency (Grant Nos. P1-0403 and N1-0232). 
\end{acknowledgments}

\section*{Authors contribution statement}
All authors listed have made a substantial, direct, and intellectual contribution to the work and approved it for publication.

\section*{Code availability}
An open-source Python module designed to efficiently calculate the $k$-nearest neighbor permutation entropy is available in GitHub at \url{github.com/hvribeiro/knnpe}. This tool can be installed from the Python Package Index (PyPI) via the command: \texttt{pip install knnpe}. 

\section*{Data availability}
The data that support the findings of this study are available from the corresponding author upon reasonable request.

\bibliography{references}

\end{document}